\documentclass{jfm}

\usepackage{graphicx}
\usepackage{natbib}

\pdfoutput=1

\ifCUPmtlplainloaded \else
  \checkfont{eurm10}
  \iffontfound
    \IfFileExists{upmath.sty}
      {\typeout{^^JFound AMS Euler Roman fonts on the system,
                   using the 'upmath' package.^^J}%
       \usepackage{upmath}}
      {\typeout{^^JFound AMS Euler Roman fonts on the system, but you
                   dont seem to have the}%
       \typeout{'upmath' package installed. JFM.cls can take advantage
                 of these fonts,^^Jif you use 'upmath' package.^^J}%
      }
  \else
  \fi
\fi

\ifCUPmtlplainloaded \else
  \checkfont{msam10}
  \iffontfound
    \IfFileExists{amssymb.sty}
      {\typeout{^^JFound AMS Symbol fonts on the system, using the
                'amssymb' package.^^J}%
       \usepackage{amssymb}%

      }{}
  \fi
\fi

\ifCUPmtlplainloaded \else
  \IfFileExists{amsbsy.sty}
    {\typeout{^^JFound the 'amsbsy' package on the system, using it.^^J}%
     \usepackage{amsbsy}}
    {}
\fi

\newsavebox{\astrutbox}
\sbox{\astrutbox}{\rule[-5pt]{0pt}{20pt}}

\title[Coherent structures in a nonlinear dynamo]{Coherent structures and the saturation of a nonlinear dynamo}

\author[E. L. Rempel, A. C.-L. Chian, A. Brandenburg and P. R. Mu\~noz]%
{E\ls R\ls I\ls C\ls O\ns L.\ns R\ls E\ls M\ls P\ls E\ls L$^1$%
  \thanks{Email address for correspondence: rempel@ita.br},\ns
A\ls B\ls R\ls A\ls H\ls A\ls M\ns C.\ls -L.\ns C\ls H\ls I\ls A\ls N$^2$\break
A\ls X\ls E\ls L\ns B\ls R\ls A\ls N\ls D\ls E\ls N\ls B\ls U\ls R\ls G$^{3,4}$ \and P\ls A\ls B\ls L\ls O\ns R.\ns M\ls U\ls \~N\ls O\ls Z$^1$}

\affiliation{$^1$Institute of Aeronautical Technology (ITA), World Institute
for Space Environment Research (WISER), S\~ao Jos\'e dos Campos -- SP 12228--900, Brazil\\[\affilskip]
$^2$Observatoire de Paris, LESIA, CNRS, 92190 Meudon, France\\[\affilskip]
$^3$NORDITA, KTH Royal Institute of Technology and Stockholm University, Roslagstullsbacken 23,
SE 10691 Stockholm, Sweden\\[\affilskip]
$^4$Department of Astronomy, Stockholm University, SE 10691 Stockholm, Sweden}

\pubyear{2010}
\volume{650}
\pagerange{119--126}
\date{?; revised ?; accepted ?. - To be entered by editorial office}
\begin{document}

\maketitle

\begin{abstract}
Eulerian and Lagrangian tools are used to detect coherent structures
in the velocity and magnetic fields of a mean--field dynamo, 
produced by direct numerical
simulations of the three--dimensional compressible magnetohydrodynamic
equations with an isotropic helical forcing and moderate 
Reynolds number. Two distinct stages of the dynamo are studied, the 
kinematic stage, where a seed magnetic field undergoes exponential growth,
and the saturated regime. It is shown that the Lagrangian analysis detects
structures with greater detail, besides providing information on the 
chaotic mixing properties of the flow and the magnetic fields. The traditional way of detecting
Lagrangian coherent structures using finite--time Lyapunov
exponents is compared with a recently developed method called function {\it M}. The latter
is shown to produce clearer pictures which 
readily permit the identification of hyperbolic regions in the magnetic field,
where chaotic transport/dispersion of magnetic field lines is highly enhanced.
\end{abstract}


\section{Introduction}

The description of chaotic and turbulent flows by means of embedded coherent structures
is a topic of great interest in the study of transport and mixing in fluids,
since these structures act as organizing units in the flow, defining 
attracting and repelling directions, transport barriers and regions of high or low 
dispersion of passive scalars. 
There is no standard way of defining what a coherent structure is,
but from the Eulerian point of view, they are often defined based on some measure related to vorticity.
An example is the highly popular $Q$--criterion, first introduced by \citet{hunt1988}
to identify vortex cores based on the difference between the rate of strain and vorticity.
Some other criteria define coherent structures or vortices
based on local pressure minima \citep{jeong1995} or on 
quantities involving the eigenvalues of the gradient tensor of the velocity field 
\citep{chong1990,zhou1999,chakraborty2005,varun2008}.
From a Lagrangian point of view, coherent structures are seen as material surfaces
that form the boundaries between regions of the flow with different behavior,
such as vortex surfaces. They are found by following trajectories of fluid particles,
while computing quantities such as the maximum rate of divergence of neighboring trajectories \citep{haller2001,shadden2005}
or the arc--length of the trajectory \citep{madrid2009}. Lagrangian tools are naturally suited
for unsteady flows, since they take into account the temporal variations of the vector field,
not just instantaneous snapshots. For a recent list of applications, see \citet{peacock2010}.

Most works on Lagrangian coherent structures (LCSs) have focused on hydrodynamic turbulence,
mainly in two--dimensions. A few papers have computed LCSs for three--dimensional magnetohydrodynamic (MHD) systems
in the conservative \citep{leoncini2006} and dissipative \citep{rempel2011,rempel2012} regimes.
In the aforementioned dissipative cases, only velocity field (kinetic) structures were explored. 
Here, we expand our previous results by computing the kinetic and magnetic coherent structures 
in a MHD model of mean--field dynamo.
Dynamo action consists in the amplification of magnetic field by the motion of an electrically conducting fluid,
being the mechanism responsible for the equipartition--strength magnetic fields observed in planets
and stars
\citep{axel2005}.
Initially, a weak magnetic field $\mathbf{B}$ undergoes an exponential growth in the
{\it kinematic dynamo} phase until $\mathbf{B}$ is strong enough to
impact the fluid velocity $\mathbf{u}$, and eventually the magnetic energy saturates. 
The saturation process is closely related to the suppression of Lagrangian chaos
in the velocity field; a comparison between the
chaoticity of the velocity field during the growth and saturation
phases of the dynamo has been performed in previous works
\citep{axel1995,cattaneo1996,zienicke1998}.
In this paper, the emphasis is on the detection of coherent structures and the transport
of passive scalars and magnetic field lines in the transition from the kinematic  
to the saturated phase. Eulerian structures are detected using the 
$Q$--criterion and, for the detection of LCSs, the traditional technique of finite--time 
Lyapunov exponents (FTLEs) is compared with the recently proposed function {\it M} \citep{madrid2009}.

Section \ref{sec:model} of this paper describes the dynamo model adopted. The numerical results are
presented in section \ref{sec:results}, where the Eulerian and Lagrangian coherent structures
are computed for the velocity and magnetic fields. Some conclusions are given in section \ref{sec:conclusions}.

\section{The Model}\label{sec:model}

The model is the prototype of $\alpha^2$ dynamo used by \citet{axel2001},
where a compressible isothermal gas is considered, with constant sound speed $c_s$, 
constant dynamical viscosity $\mu$, constant magnetic diffusivity 
$\eta$, and constant magnetic permeability $\mu_0$. 
The following set of compressible MHD equations is solved 

\begin{eqnarray}
& &\partial_t \ln\rho+\mathbf{u}\cdot\nabla\ln\rho+\nabla\cdot\mathbf{u}=0,\label{eq continuity}\\
& &\partial_t\mathbf{u}+\mathbf{u\cdot}\nabla\mathbf{u}=-\nabla p / \rho + \mathbf{J\times B}/\rho+ 
    (\mu/\rho)\left(\nabla^{2}\mathbf{u}+\nabla\nabla\cdot\mathbf{u}/3\right)+\mathbf{f},\label{eq momentum}\\
& &\partial_t\mathbf{A}=\mathbf{u\times B}-\eta\mu_{0}\mathbf{J}\label{eq induction},
\end{eqnarray}
where $\rho$ is the density, $\mathbf{u}$ is the fluid velocity, $\mathbf{A}$ is the magnetic vector potential,
 $\mathbf{J} = \nabla \times \mathbf{B}/\mu_0 $ is the current density, $p$ is the 
pressure,  
$\mathbf{f}$ is an external forcing, and $\nabla p / \rho = c_s^2\nabla\ln\rho$, 
where  $c_s^2 = \gamma p/\rho$ is assumed to be constant. 
Nondimensional units are adopted by setting $k_1=c_s = \rho_0 = \mu_0 = 1$, where $\rho_0=\left<\rho\right>$ 
is the spatial average of $\rho$
and $k_1$ is the smallest wavenumber in the box, which has sides $L=2\pi$ and periodic boundary conditions.
Thus, the time unit is $(c_sk_1)^{-1}$, space is measured in units of $k_1^{-1}$, $\mathbf{u}$ in units of $c_s$,
$\mathbf{B}$ in units of $(\mu_0\rho_0)^{1/2}c_s$, $\rho$ in units of $\rho_0$
and the unit of viscosity $\nu \equiv \mu/\rho_0$ and magnetic diffusivity $\eta$ is $c_s/k_1$. 
Equations (\ref{eq continuity})--(\ref{eq induction}) are solved with the 
{\small PENCIL CODE} \footnote{http://pencil-code.googlecode.com}, which
employes an explicit sixth--order finite differences scheme in space and a third--order Runge--Kutta scheme for time integration.

The initial conditions are $\ln \rho = \mathbf{u} = 0$, 
and $A$ is a set of normally distributed, uncorrelated random numbers with zero mean
and standard deviation 
equal to $10^{-3}$. The forcing function $\mathbf{f}$ is given by

\begin{equation}
\mathbf{f}(\mathbf{x},t)=\mbox{Re}\{ N\mathbf{f}_{\mathbf{k}(t)}\exp[i\mathbf{k}(t)\cdot\mathbf{x}+i\phi(t)]\},
\end{equation}
where $\mathbf{k}(t) = (k_x , k_y , k_z)$ is a time--dependent wavevector, $\mathbf{x} = (x, y, z)$
 is position, and
$\phi(t)$ with $|\phi| < \pi$ is a random phase. On dimensional grounds the normalization factor
is chosen to be $N = f_0 c_s (kc_s /\delta t)^{1/2}$, where $f_0$ is a nondimensional factor, 
$k = |\mathbf{k}|$, and $\delta t$
is the length of the integration timestep. We focus on the case where
$|\mathbf{k}|$ is around $k_f=5$ and randomly select, at each timestep, one of 350 possible vectors in
$4.5 < |\mathbf{k}| < 5.5$. The operator $\mathbf{f_k}$ is given by

\begin{equation}
\mathbf{f}_\mathbf{k}=\frac{i\mathbf{k} \times (\mathbf{k} \times \mathbf{e}) - |\mathbf{k}|(\mathbf{k} \times \mathbf{e})}
{\mathbf{k}^2\sqrt{2(1-(\mathbf{k\cdot e})^2)/\mathbf{k}^2}},
\end{equation}
where $\mathbf{e}$ is an arbitrary unit vector needed in order to generate a vector $\mathbf{k \times e}$ that is
perpendicular to $\mathbf{k}$. Note that $|\mathbf{f_k}|^2 = 1$ and the helicity density satisfies 
$\mathbf{f \cdot \nabla \times f=|k|f^2>0}$, which is an important condition for 
the production of a mean--field dynamo \citep{moffatt1978}. 
The forcing function is delta--correlated in time, i. e., all points of $\mathbf{f}$ are correlated
at any instant in time but are different at the next time step.
Following \citet{axel2001}, the control parameters are set as $f_0 = 0.07$, $\nu=\eta=0.002$ and
the numerical resolution is $128^3$.

\section{Results}\label{sec:results}

\subsection{Mean--field dynamo}\label{sec:dynamo}

\begin{figure}
  \centerline{\includegraphics[width=0.6\columnwidth]{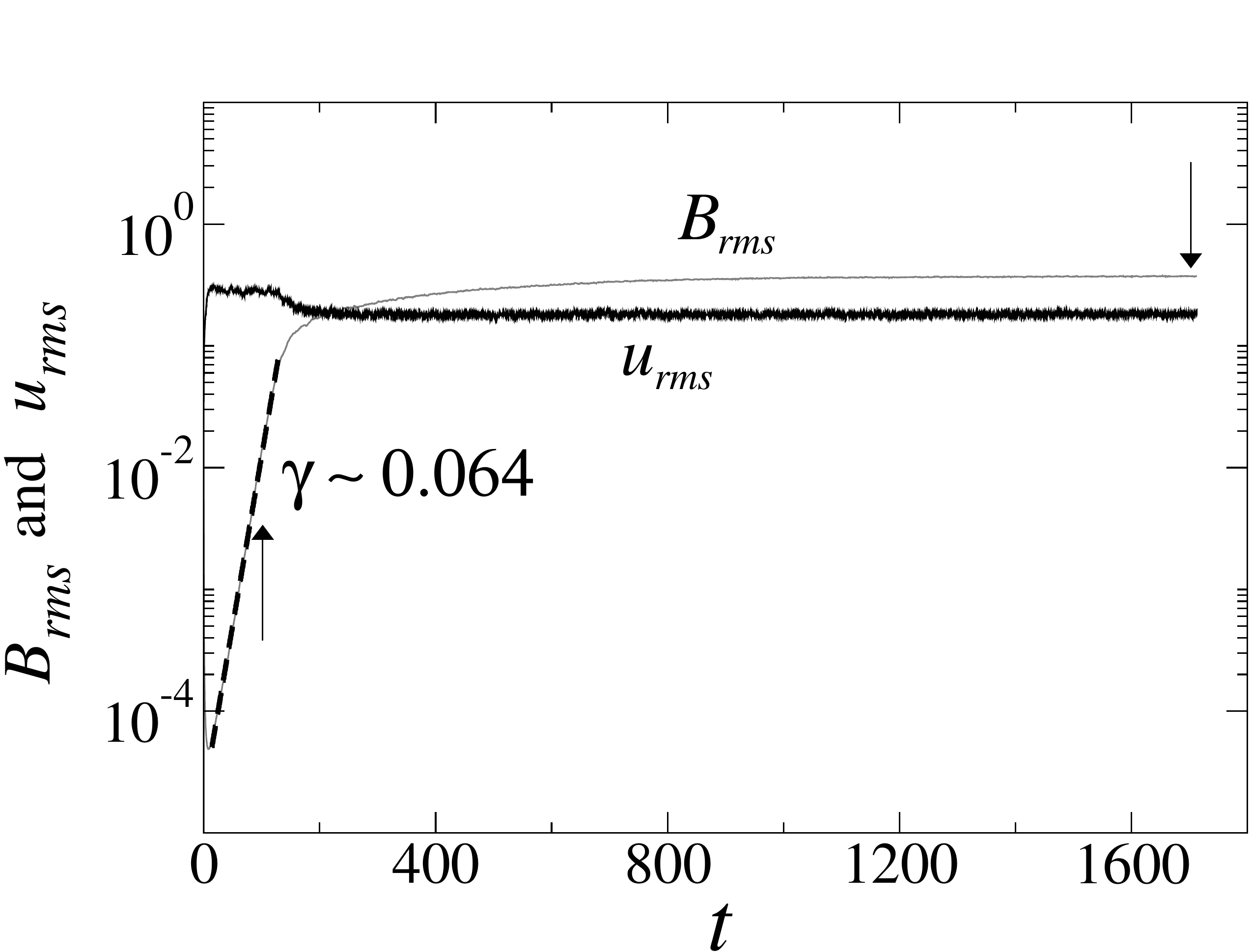}}
  \caption{Time series of $B_{rms}$ (light line) and $u_{rms}$ (dark line) of MHD dynamo simulations for $\eta=\nu=0.001$.
The arrows indicate the kinematic phase at $t=100$ and the saturated nonlinear regime at $t=1700$, respectively. The growth
rate during the kinematic phase is $\gamma \sim 0.064$.}
\label{fig ts}
\end{figure}

Figure \ref{fig ts} shows the time series of $B_{rms}\equiv \left< B^2 \right>^{1/2}$ (light line) and $u_{rms}\equiv \left< u^2 \right>^{1/2}$ 
(dark line),
where $B\equiv |\mathbf{B}|$ and $u\equiv |\mathbf{u}|$. During the first time units up to $t\sim 150$, the magnetic energy is too week to
significantly impact the velocity field and $u_{rms} \sim 0.28$, thus the Reynolds number is $Re=u_{rms}/\nu k_f \sim 28$. 
During this {\it kinematic phase}, $B_{rms}$ increases exponentially,
with a growth rate $\gamma \sim 0.064 \pm 2\times 10^{-5}$ obtained from the fitted line (dashed line). After $t\sim 150$, $u_{rms}$ starts to
decay due to the contribution of the Lorentz force (second term in the right side of Eq. (\ref{eq momentum})). 
Eventually, the {\it r.m.s.} quantities saturate due to nonlinear 
effects, with $u_{rms} \sim 0.18$ while the magnetic field reaches a super--equipartition value $B_{rms} \sim {0.37} > u_{rms}$.
The arrows indicate the times $t=100$ and $t=1700$, respectively, which will be used later to represent the kinematic and saturated phases.
In turnover time units  ($1/ k_fu_{rms}$), the referred times are  $u_{rms} k_f t \sim 140$ and $u_{rms} k_f t \sim 1530$, respectively,
and the growth rate is $\gamma/u_{rms}k_f \sim 0.046$.

\begin{figure}
  \centerline{\includegraphics[width=1.\columnwidth]{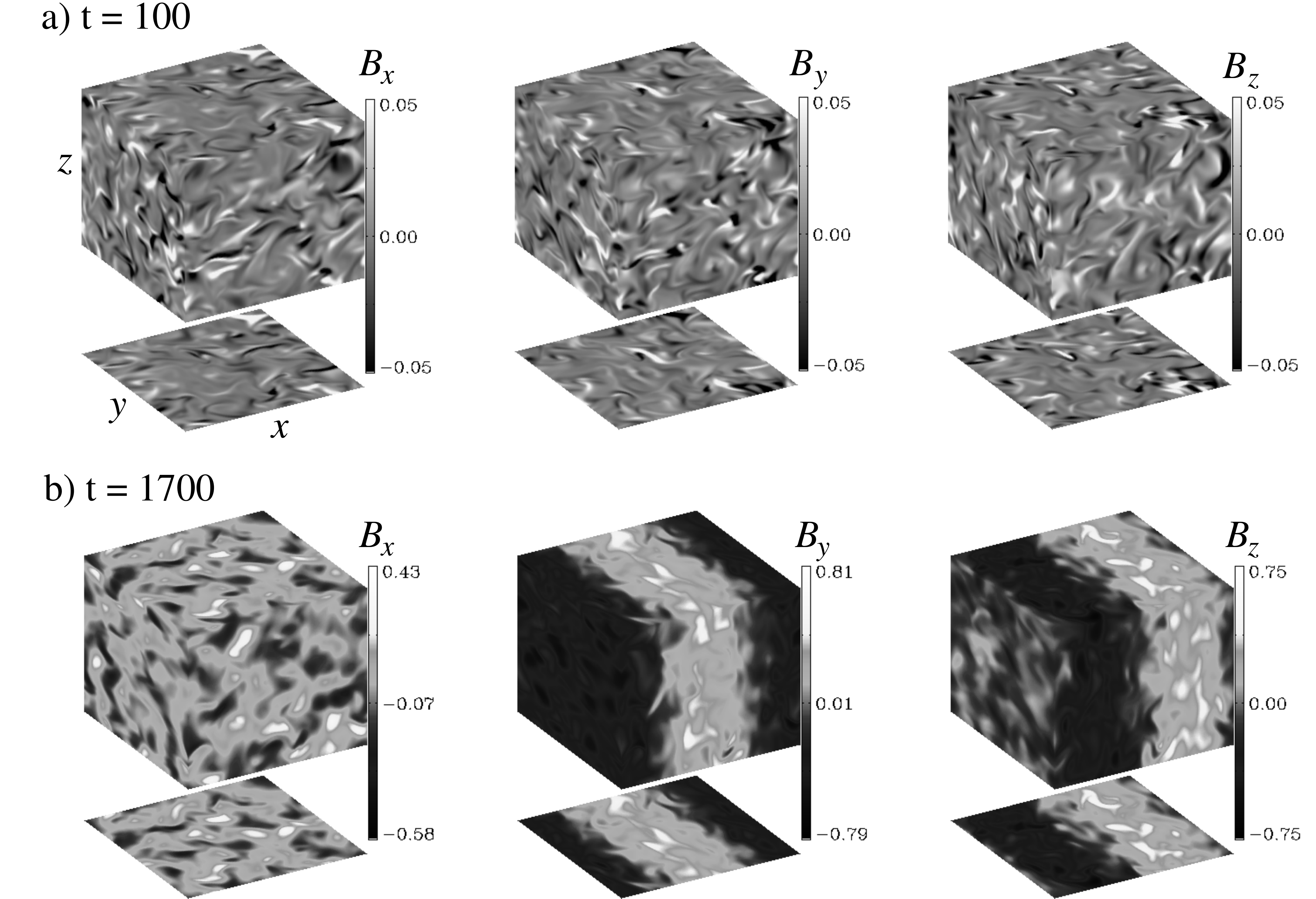}}
  \caption{Intensity plot of magnetic field components at $t=100$ (upper panel) and $t=1700$ (lower panel).}
\label{fig bs}
\end{figure}

During the kinematic stage, the magnetic field displays low--amplitude stochastic fluctuations, as shown in 
the upper panels of Fig. \ref{fig bs}. As $B_{rms}$ grows, small--scale velocity and magnetic field fluctuations 
combine to produce a robust large--scale mean--field pattern (lower panels). The physics behind the rise of this mean--field
is related to the so--called $\alpha$--effect \citep{moffatt1978} and has been explored 
in this model by \citet{axel2001}.
  
\subsection{Eulerian coherent structures}\label{sec:euler}

Eulerian coherent structures can be extracted from the velocity field by decomposing the gradient tensor $\nabla \mathbf{u}$
as

\begin{equation}
\mathbf{\nabla} \mathbf{u} = \mathbf{S} + \mathbf{\Omega},
\end{equation}
where $\mathbf{S}=\frac{1}{2}[\nabla\mathbf{u}+\left(\nabla \mathbf{u}\right)^T]$
and $\mathbf{\Omega}=\frac{1}{2}[\nabla\mathbf{u}-\left(\nabla \mathbf{u}\right)^T]$
are the symmetric and antisymmetric parts of $\nabla \mathbf{u}$, respectively. The symmetric part is 
the rate--of--strain tensor and the antisymmetric part is the vorticity tensor. One way to define
an Eulerian coherent structure is by finding regions of $\mathbf{u}$ where vorticity dominates over strain,
which can be measured by the $Q$--criterion \citep{hunt1988,zhong1998,haller2005,lawson2010}

\begin{equation}
Q=\frac{1}{2}[|\mathbf{\Omega}|^2-|\mathbf{S}|^2].
\end{equation}
Thus, an Eulerian coherent structure or vortex is defined as a region where $Q>0$.

\begin{figure}
  \centerline{\includegraphics[width=1.\columnwidth]{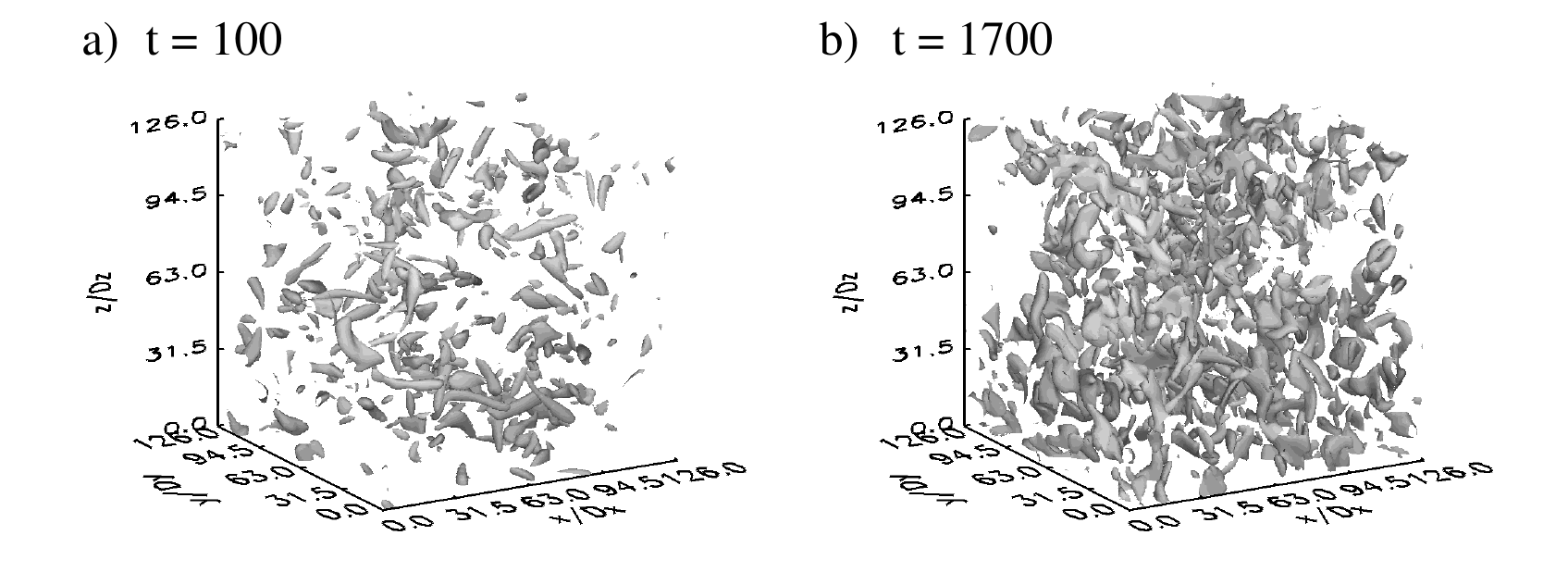}}
  \caption{Eulerian coherent structures in the velocity field, detected by instantaneous
           isosurfaces of the $Q$--criterion.
           The isosurfaces are defined using $15\%$ maximum $Q$.}
\label{fig qu3d}
\end{figure}

\begin{figure}
  \centerline{\includegraphics[width=1.\columnwidth]{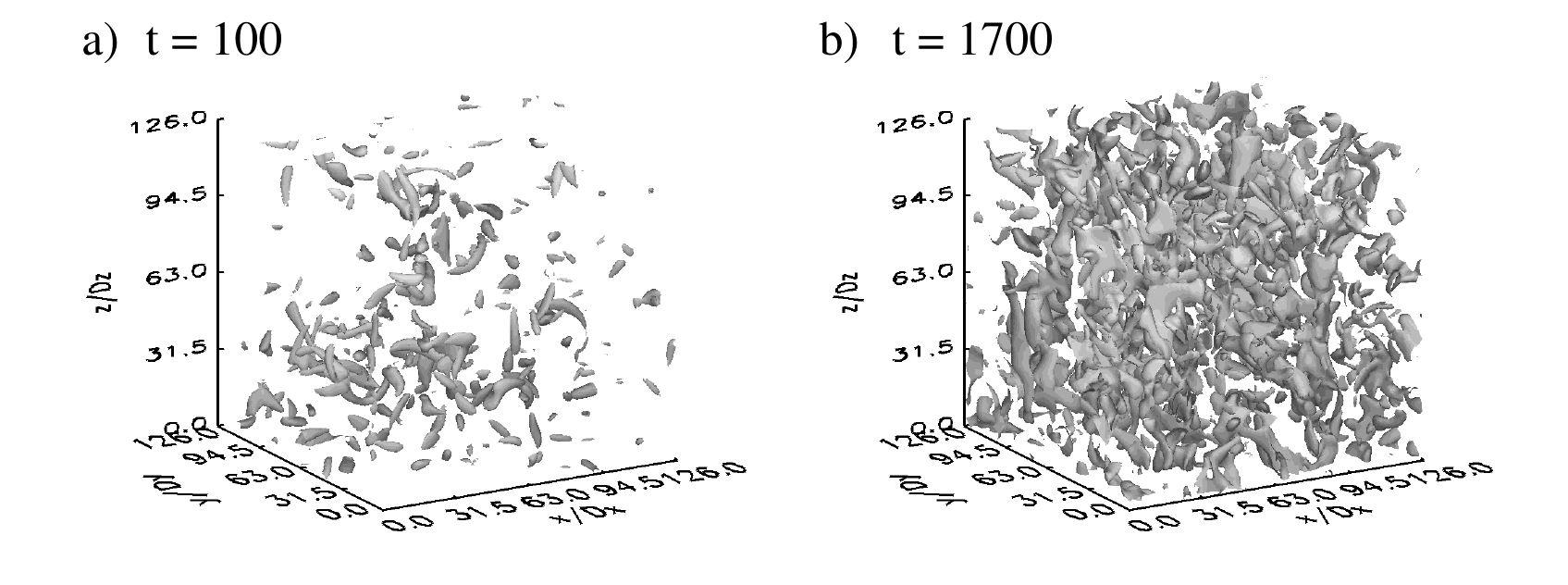}}
  \caption{Eulerian coherent structures in the magnetic field, detected by
           isosurfaces of the $Q$--criterion. The isosurfaces are defined using $15\%$ maximum $Q$.}
\label{fig qb3d}
\end{figure}

Figure \ref{fig qu3d} shows the isosurfaces of the $Q$--criterion, using $15\%$ maximum $Q$ (contour surfaces enclose
high $Q$ values). These plots are highly dependent on the threshold chosen for $Q$,
but it is possible to see that the fluid is more intermittent at the kinematic dynamo phase ($t=100$)
than after saturation ($t=1700$), since in the right panel the coherent structures fill the space
in a more homogeneous way. There are fewer regions for $t=100$ where $Q$ is much higher than the average,
thus the presence of fewer vortices for this threshold in Fig. \ref{fig qu3d}(a) than in \ref{fig qu3d}(b), where
local values of $Q$ are closer to the average $Q$. Figure \ref{fig qb3d} shows the corresponding 
plots of $Q$ for the magnetic field, where the coherent structures represent magnetic vortices or current structures
\cite{axel1996}.

\begin{figure}
  \centerline{\includegraphics[width=1.\columnwidth]{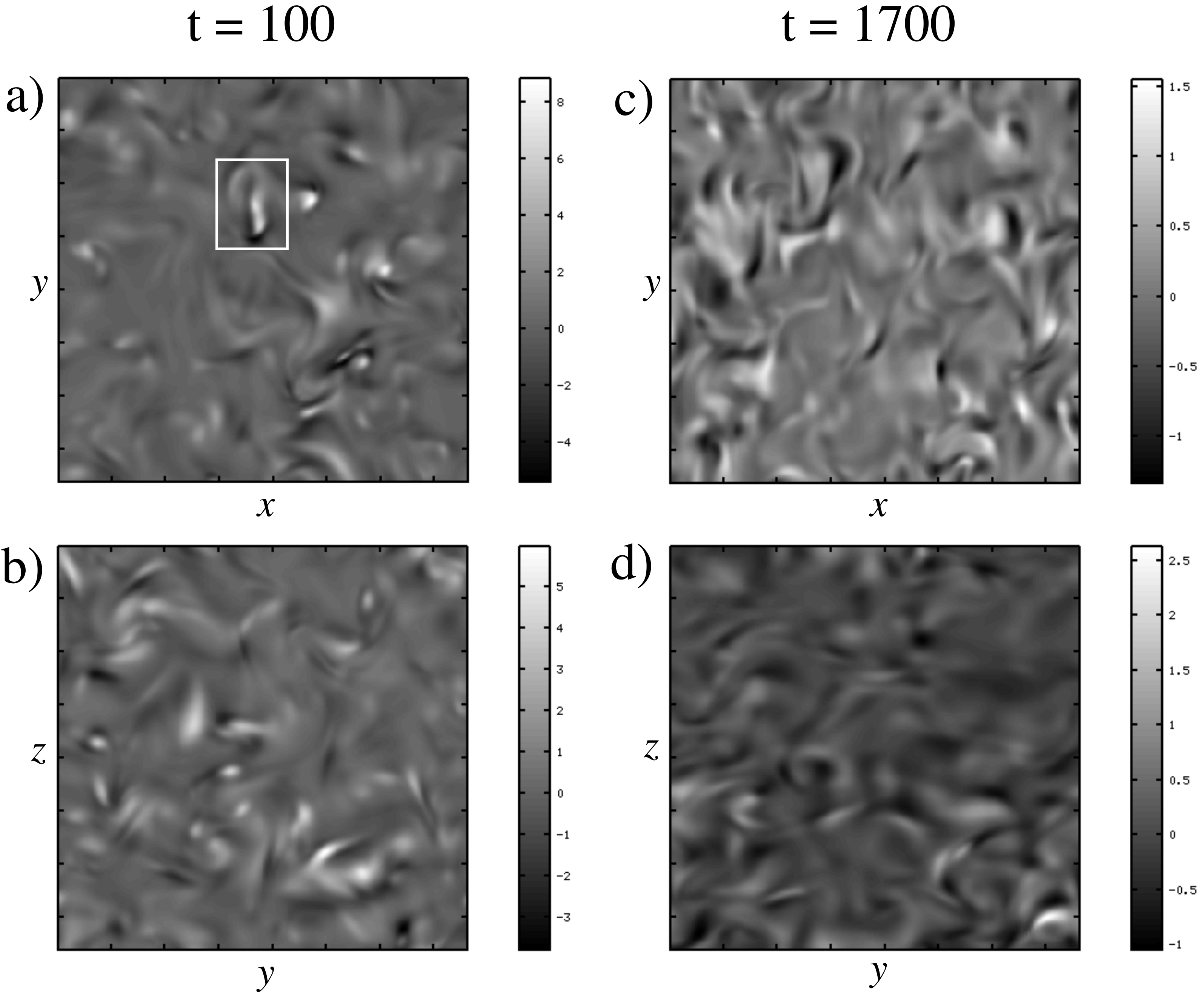}}
  \caption{Eulerian coherent structures in the velocity field, detected by the 
           $Q$--criterion at $t=100$ (left panel) and $t=1700$ (right panel).}
\label{fig qu}
\end{figure}

\begin{figure}
  \centerline{\includegraphics[width=1.\columnwidth]{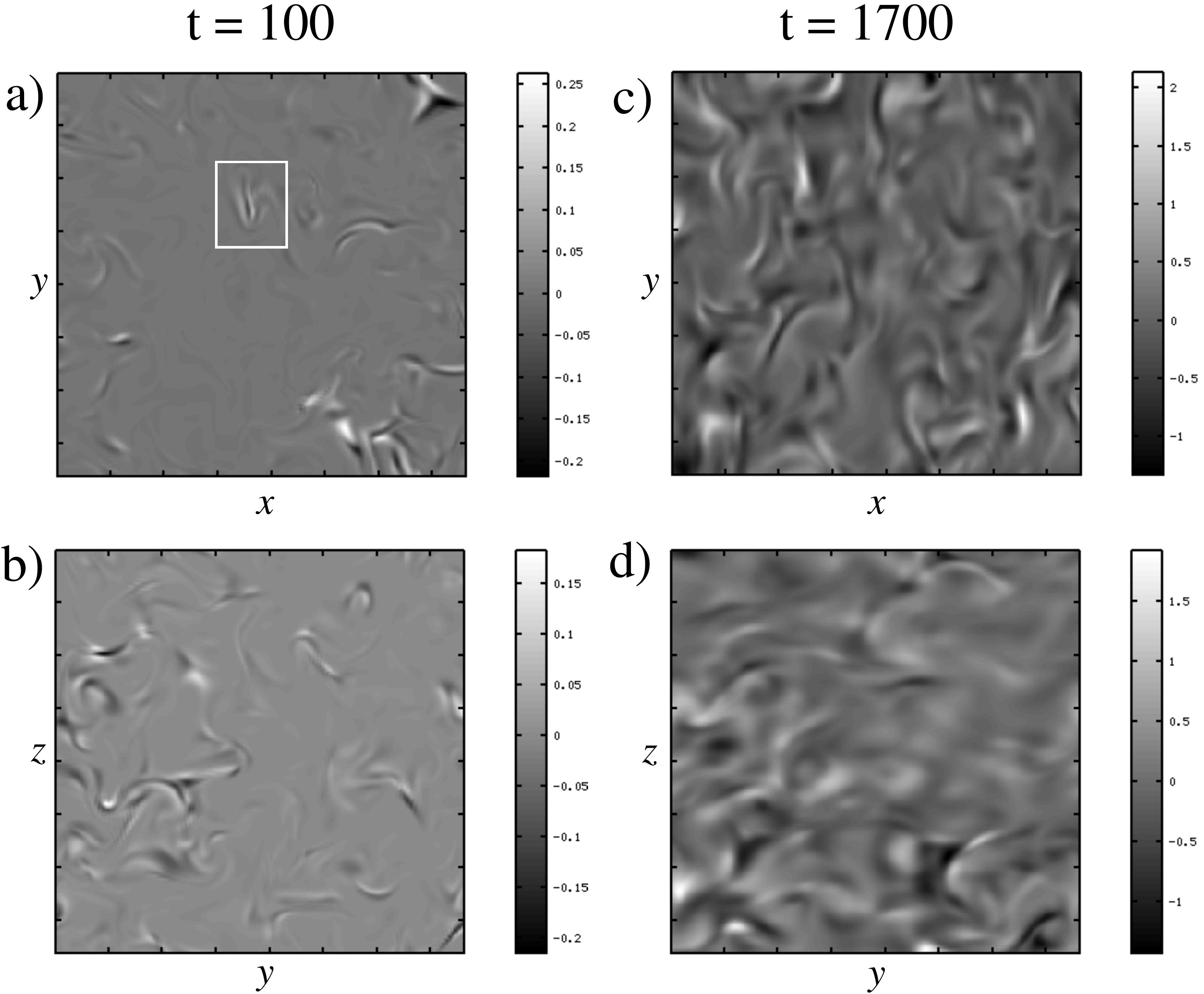}}
  \caption{Eulerian coherent structures in the magnetic field, detected by the 
           $Q$--criterion at $t=100$ (left panel) and $t=1700$ (right panel).}
\label{fig qb}
\end{figure}

In Fig. \ref{fig qu}, intensity plots of the $Q$--criterion are shown for two--dimensional 
slices of the box at planes $z=0$ (upper panels) and $x=0$ (lower panels) at times $t=100$ (left panels)
and $t=1700$ (right panels), respectively. Coherent structures with strong vorticity are observed as bright spots,
such as the one highlighted by a box in Fig. \ref{fig qu}(a). 
Notice that at $t=1700$ a large number of bright spots is seen in the $xy$--plane, but they are rare in the
$yz$--plane, revealing a preferential alignment of coherent structures in the vertical direction
in the saturated regime.
A similar plot is shown for the magnetic field in Fig. \ref{fig qb},
where some of the same coherent structures found in the velocity field can be observed,
reflecting the strong coupling between $\mathbf{B}$ and $\mathbf{u}$ in 
Eqs. (\ref{eq momentum}) and (\ref{eq induction}).

Although some coherent structures are clearly detected by this Eulerian technique, 
the $Q$--criterion relies on a user defined threshold to determine their boundaries.
In order to precisely identify the boundaries and the main transport barriers in the flow,
the next section proceeds with a Lagrangian analysis.

\subsection{Lagrangian coherent structures}\label{sec:lcs}

This section describes two tools that can be employed to
define/detect coherent structures in the Lagrangian frame,
the finite--time Lyapunov exponents and the recently proposed function {\it M}.

\subsubsection{Finite--time Lyapunov exponents}
\label{sec lyap}

In the Lagrangian point of view, coherent structures
are seen as material surfaces around which trajectory patterns
are formed. In \citet{haller2000}, these surfaces are simply called
Lagrangian coherent structures and are distinguished
from other material surfaces in that a LCS exhibits locally
the strongest attraction, repulsion or shearing in the flow.
Repelling LCSs are responsible for generating stretching, 
attracting LCSs for folding, and shear LCSs for swirling and 
jet-type tracer patterns \citep{haller2011}.

Attracting LCSs have commonly been associated with local maximizing curves ({\it ridges})
in the backward--time finite--time Lyapunov exponent (FTLE) field and repelling LCSs to ridges in the forward--time
FTLE field \citep{shadden2005,green2007,beron2010}. There are limitations in such definition,
as pointed out by \citet{haller2011} and \citet{haller2012}, e. g., a ridge in the FTLE field may indicate 
the presence of a shear LCS or no LCS at all. But in general, 
ridges in the FTLE fields provide a good approximation to the true LCSs of the flow.

Let $D\subset\mathbb{R}^{3}$ be the domain of the fluid to be studied,
let $\mathbf{x}(t_{0})\in D$ denote the position of a passive particle
at time $t_{0}$ and let $\mathbf{u}(\mathbf{x},t)$ be the velocity
field defined on $D$. The motion of the particle is given by the
solution of the initial value problem 

\begin{equation}
\frac{d\mathbf{x}}{dt}=\mathbf{u}(\mathbf{x}(t),t), \qquad \mathbf{x}(t_{0})=\mathbf{x}_{0}.
\label{eq odes}
\end{equation}

Let the flow map for $\mathbf{u}$ be defined as $\phi_{t_{0}}^{t_{0}+\tau}:\mathbf{x}(t_{0})\mapsto\mathbf{x}(t_{0}+\tau)$.
The deformation gradient is given by $J=\mathrm{d}\phi_{t_{0}}^{t_{0}+\tau}(\mathbf{x})/\mathrm{d}\mathbf{x}$
and the finite-time right Cauchy-Green deformation tensor is given
by $\triangle=J^{T}J$. Let $\lambda_{1}>\lambda_{2}>\lambda_{3}$
be the eigenvalues of $\triangle$. Then, the finite-time Lyapunov
exponents or direct Lyapunov exponents of the trajectory of
the particle are defined as

\begin{equation}
\sigma_{i}^{t_{0}+\tau}(\mathbf{x})=\frac{1}{|\tau|}\ln\sqrt{\lambda_{i}},\qquad i=1,2,3.
\label{eq ftle}
\end{equation}

A positive $\sigma_{1}$ is the signature of chaotic streamlines in
the velocity field, being a measure of the stretching of fluid elements 
(although it also incorporates shear \citep{haller2011}). In this work, the deformation
gradient is computed with second order centered finite--differences.

\subsubsection{Function {\it M}}

\citet{madrid2009} proposed a function
to define ``distinguished trajectories" (DTs), which are a generalization
of the concept of fixed points for aperiodically time--dependent flows.
In stationary flows, hyperbolic fixed points are responsible for particle
dispersion and nonhyperbolic fixed points for particle confinement.
Invariant stable and unstable manifolds of hyperbolic fixed points
are barriers to transport and divide the phase space in regions with qualitatively different behaviors.
The proposed function, named {\it function M}, can reveal
both hyperbolic and nonhyperbolic flow regions of time--dependent flows. Moreover, {\it M} is
also useful to detect the stable and unstable manifolds of 
distinguished hyperbolic trajectories (DHTs), defined as the set of points
such that trajectories passing through these points at $t=t_0$ will
approach the DHTs at an exponential rate as time goes to infinity or
minus infinity, respectively \citep{branicki2011}. The stable and unstable
manifolds of DHTs correspond to the repelling and attracting Lagrangian coherent
structures, respectively, as defined in the section \ref{sec lyap}.

Consider the system given by Eq. (\ref{eq odes}), where $\mathbf{x}=\{x_1,x_2,x_3\}$.
For all initial conditions $\mathbf{x}_0$ in $D$
at a given time $t_0$, let us define the function $M(\mathbf{x}_0,t_0):(D,t)\rightarrow \mathbb{R}$
as

\begin{equation}
M(\mathbf{x}_0,t_0)_\tau = \int_{t_0-\tau}^{t_0+\tau}\left(\sum_{i=1}^3(dx_i(t)/dt)^2\right)^{1/2}dt.
\label{eq m}
\end{equation}
Thus, the function {\it M} is a measure of the arc length of the curve traced by $\mathbf{x}_0$. 
Local minima of {\it M} represent trajectories that ``move less", being
related either to hyperbolic or nonhyperbolic DTs.
The manifolds of DHTs are also visible in the {\it M} field, since
one expects a sharp distinction in the lengths of trajectory curves 
for particles in regions with different behaviors, separated by stable and unstable manifolds,
as noted by \citet{mendoza2010}. The technique has been successfully applied
to the detection of DTs and manifolds in oceanic \citep{mendoza2010,mendoza2010b} and
stratospheric \citep{camara2012} flows.

\subsubsection{Velocity field structures and chaotic mixing}\label{sec:vel}

\begin{figure}
  \centerline{\includegraphics[width=1.\columnwidth]{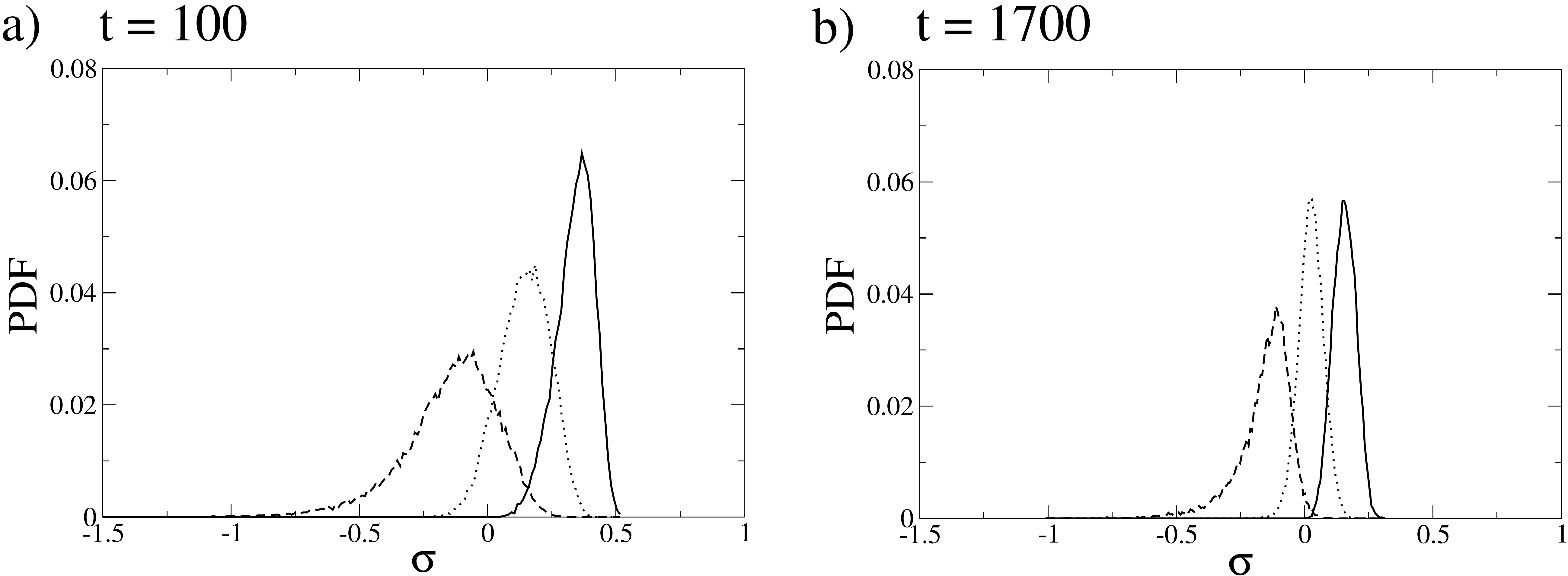}}
  \caption{Probability density functions (PDFs) of the FTLEs of the velocity field
           at the kinematic ($t=100$, left panel) and saturated ($t=1700$, right panel) regimes.
           The solid line represents $\sigma_1$, the dotted line, $\sigma_2$, and the dashed line, $\sigma_3$.}
\label{fig pdfs}
\end{figure}

The FTLEs are computed from a series of fully 3D snapshots
of the velocity field taken at different times from $t_0$ to $t_0 + \tau$.
Linear interpolation in time and third-order splines in
space are used to obtain the continuous vector fields necessary
to obtain the particle trajectories. Figure \ref{fig pdfs} depicts the probability distribution functions (PDFs) of the 
three FTLEs at $t_0=100$ (left) and $t_0=1700$ (right) computed for $64^3$ particle trajectories
from Eq. (\ref{eq ftle}) with a value of $\tau$ corresponding to 9 turnover time
units, where $u_{rms}\sim 0.28$ for the kinematic phase and
$u_{rms} \sim 0.18$ for the saturated regime. Therefore, $\tau=9/(k_fu_{rms}) \sim 6.4$ time units 
for the kinematic phase and $\tau \sim 10$ time units for the saturated phase.
One can see a clear reduction of Lagrangian chaos in the velocity
field at $t_0=1700$, with the PDF of $\sigma_1$ being shrunk and shifted to the left. There are also 
fewer regions with two or three positive exponents. Overall, chaotic mixing is diminished due to 
the growth of $B_{rms}$ and the action of the Lorentz force. The asymmetry in the distributions is typical 
of heterogeneous mixing, where both regular and irregular trajectories coexist (in finite--time), 
which means that trajectories cannot uniformly sample the
phase space (see, e. g., \citet{beron2010}).

From Fig. \ref{fig pdfs} it is clear that most trajectories display
two positive Lyapunov exponents. \citet{zeldovich1984} and \citet{chertkov1999} state that in such a case, the total
magnetic energy in a kinematic dynamo should behave as $B^2 \propto \exp[(\sigma_1 - \sigma_2)t]$,
therefore, one has for the growth rate

\begin{equation}
\gamma = \frac{d\ln\left< B^2 \right>^{1/2}}{dt} = \frac{d\ln \left\{ \exp[(\sigma_1 -\sigma_2)t/2]\right\} }{dt}
       = \frac{\sigma_1 - \sigma_2}{2}.
\label{eq gamma}
\end{equation}
At $t_0=100$, $\left< \sigma_1 \right> \sim 0.339$ and $\left< \sigma_2 \right> \sim 0.143$, which 
from Eq. (\ref{eq gamma}) provides $\gamma = 0.098$ (or $\gamma = 0.07$ in dimensional units), 
which agrees to within an order of magnitude with the fitted value
$\gamma \sim 0.064$, given in Fig. \ref{fig ts}. 
 
\begin{figure}
  \centerline{\includegraphics[width=1.\columnwidth]{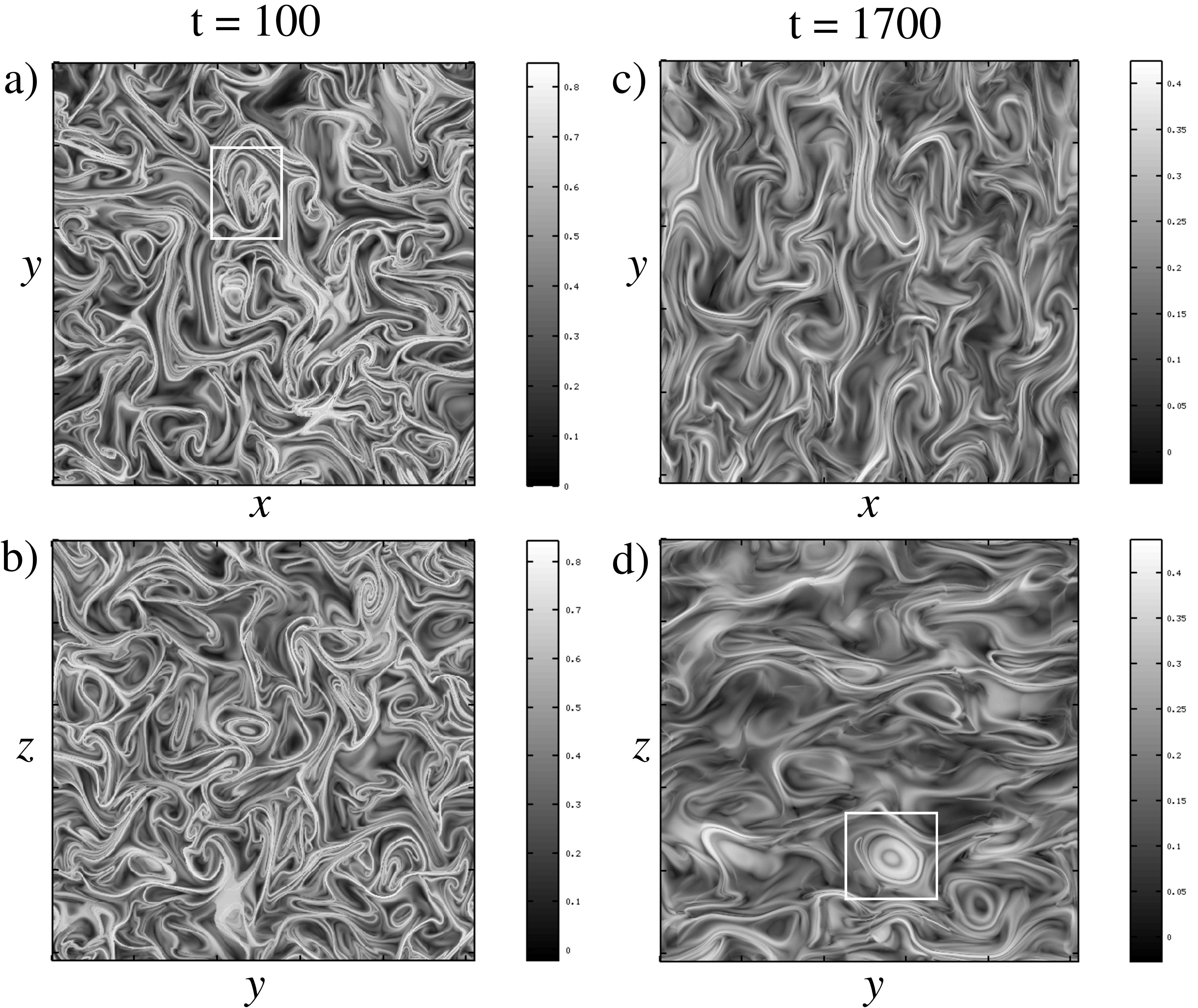}}
  \caption{Attracting Lagrangian coherent structures in the velocity field, 
           given by the backward--time FTLE at $t_0=100$ (left panel) and $t_0=1700$ (right panel).}
\label{fig umu}
\end{figure}

The remainder of this paper focuses on the backward--time maximum FTLE field, since
they reveal the attracting LCSs, which correspond to structures seen using flow visualization in experiments \citep{green2007}.
Figure \ref{fig umu} shows the backward--time maximum FTLE field computed for $\tau$ corresponding to 9 turnover time units
at $t_0=100$ (left) and $t_0=1700$ (right) from a grid of initial conditions 
with $512\times 512$ particles. The bright lines represent the attracting LCSs.
While the LCSs at $t_0=100$ reveal no preferred direction, consistent with an isotropic forcing,
at $t_0=1700$ there is a clear vertical alignment of LCSs in the $xy$--slice (Fig. \ref{fig umu}(b)).
This is due to the super--equipartition magnetic field at $t_0=1700$, which develops a large--scale
vertical pattern in this plane (see Fig. \ref{fig bs}), affecting the alignment of velocity field vectors. 

A comparison between Figs. \ref{fig qu} and \ref{fig umu} shows that the FTLE field provides a clearer depiction of
coherent structures, with finer details and more precise detection of structure boundaries.
Moreover, some coherent structures are only apparent in the FTLE field, such as the large eddy 
indicated in Fig. \ref{fig umu}(d).

\begin{figure}
  \centerline{\includegraphics[width=1.\columnwidth]{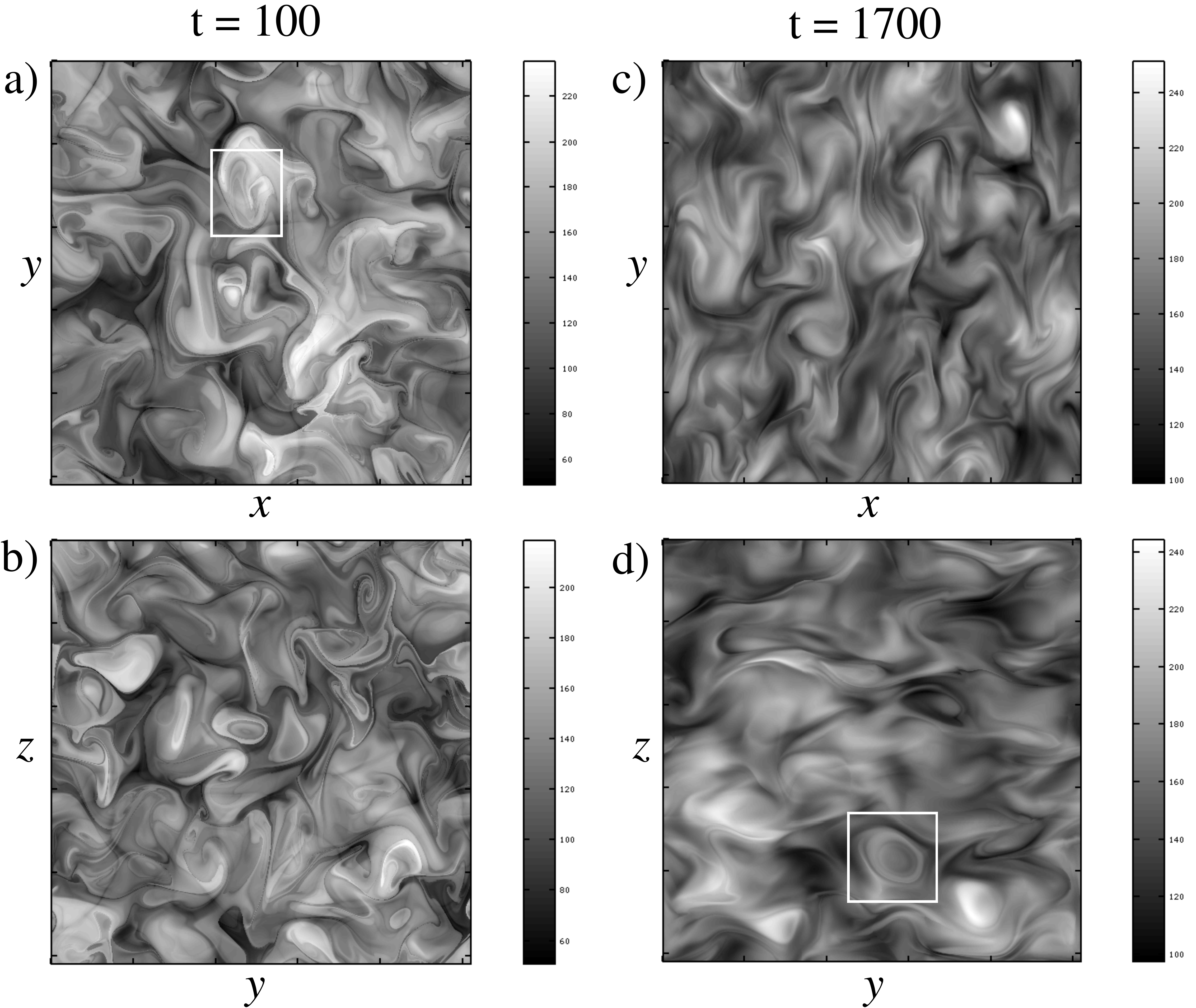}}
  \caption{Lagrangian coherent structures in the velocity field, given by 
           the function $M$ at $t_0=100$ (left panel) and $t_0=1700$ (right panel).}
\label{fig mu}
\end{figure}

From our experience, one of the problems with FTLE plots in turbulent flows 
is that pictures usually become increasingly complex for larger $\tau$,
with material lines ``growing" and filling the entire phase space. In that sense, it is easier to use function {\it M}
to detect the main coherent structures of the flow. Figure \ref{fig mu} is a plot of function {\it M}
with $\tau=9$ in turnover time units, equivalent to Fig. \ref{fig umu}. For the kinematic phase at $t_0=100$ (left panel)
the eddies are clearly identified as closed regions with distinguished shades of gray. At $t_0=1700$ (right panel)
the borders between regions are not so sharp and there are wide smooth regions in the flow.
Smoothness in the {\it M} field indicates that trajectories in those regions 
do not reach nearby hyperbolic regions during $\pm\tau$ turnover time units, since hyperbolic
trajectories are the ones responsible for dispersion and for producing sharp changes in M \citep{mendoza2010}.
For larger $\tau$, the boundaries become sharper and more foldings of manifolds are seen, but we
keep $\tau=9$ in all our pictures to facilitate the comparison between both methods in different regimes.
Overall, the function {\it M} seems to be less sensitive  to the choice of $\tau$ than the FTLE. 
Figure \ref{fig mu} corroborates with Fig. \ref{fig pdfs} in revealing that there is less
chaotic mixing after the nonlinear saturation of $B_{rms}$.

\subsubsection{Magnetic field structures and transport of field lines}\label{sec:mag}

Our simulations reveal that the magnetic field displays smooth and complex regions (see Fig. \ref{fig qb}).
If one applies the Lagrangian techniques discussed in the previous section to the magnetic field,
the identification of magnetic LCSs provides the main barriers to the transport
of field lines, a topic of great interest in magnetic reconnection studies \citep{evans2004,grasso2010,borgogno2011,yeates2011}.


To obtain the magnetic LCSs, the magnetic field at a fixed dynamic time $t_0$ is used and 
the maximum FTLE field is computed by integrating 

\begin{equation}
\frac{d\mathbf{x}}{ds}=\mathbf{B}(\mathbf{x}(s),t_0), \qquad \mathbf{x}(s_0)=\mathbf{x_0},
\label{eq line}
\end{equation}
where the parameter (position) $s$ along the field line is seen as an effective time,
or field--line--time \citep{borgogno2011}. The flow map for $\mathbf{B}$ is defined 
as $\phi_{s_0}^{s_0+\tau}:\mathbf{x}(s_0)\rightarrow \mathbf{x}(s_0+\tau)$. 
Equation (\ref{eq line}) is integrated from $s_0$ to $s_0 + \tau$ with $t$
fixed at $t_0$.
Lagrangian chaos in the magnetic field is 
responsible for the transport of magnetic field lines between different regions of the box.
Here, the term ``transport" is used to refer to motion of field lines in field--line--time,
not in dynamic time. Therefore, the maximum FTLE provides a measure of the exponential separation
between two neighboring field lines after a finite field--line--time $\tau$, i. e., after a finite distance 
along the field line. 

\begin{figure}
  \centerline{\includegraphics[width=1.\columnwidth]{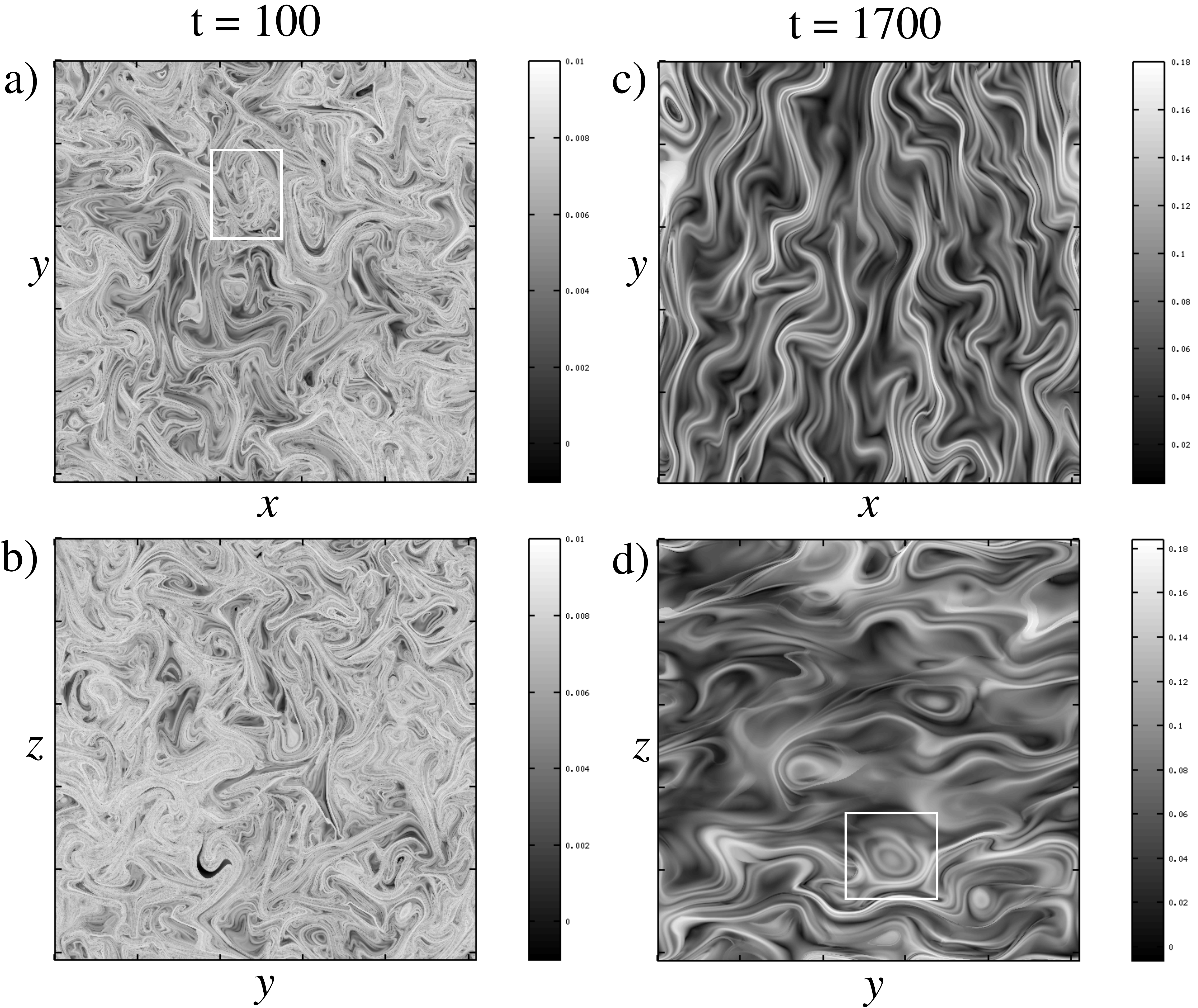}}
  \caption{Attracting Lagrangian coherent structures in the magnetic field, given by 
           the backward--time maximum FTLE at $t=100$ (left panel) and $t=1700$ (right panel).}
\label{fig umB}
\end{figure}

Figure \ref{fig umB} shows the backward--time maximum FTLE field for the kinematic (left panel) and saturated (right panel) regimes.
The high--intensity lines represent attracting magnetic LCSs which act as barriers to field line transport. 
No transport of magnetic field lines occurs across invariant LCSs and large--scale 
transport is possible only through homoclinic and heteroclinic crossings of attracting and repelling LCSs, where a lobe 
dynamics mechanism takes place \citep{grasso2010,borgogno2011,yeates2011,rempel2012}.
Both FTLE fields are obtained by fixing the evolution (dynamic) time ($t_0=100$ for the left panel and $t_0=1700$ for the right panel) 
and setting $\tau = 9/B_{rms}$, where $B_{rms}=0.014$ for $t_0=100$
and $B_{rms}=0.37$ for $t_0=1700$. In the kinematic regime ($t_0=100$) the LCSs display no 
preferred direction, and randomly fill the simulation box. 
Note that, at least for this
value of $\tau$, it is difficult to identify the coherent structure marked in the box in Fig. \ref{fig umB}(a)
due to the many foldings of attracting lines.
After growth and saturation of $B_{rms}$ ($t_0=1700$), the randomness of field line orientation is diminished 
and there is a preferential direction of alignment of field lines which, as mentioned before, directly affects the velocity field.
The growth of $B_{rms}$ is also reflected in the presence of many smooth regions in Figs. \ref{fig umB}(c) and (d).
Consequently, there is less chaotic mixing of field lines. 

\begin{figure}
  \centerline{\includegraphics[width=1.\columnwidth]{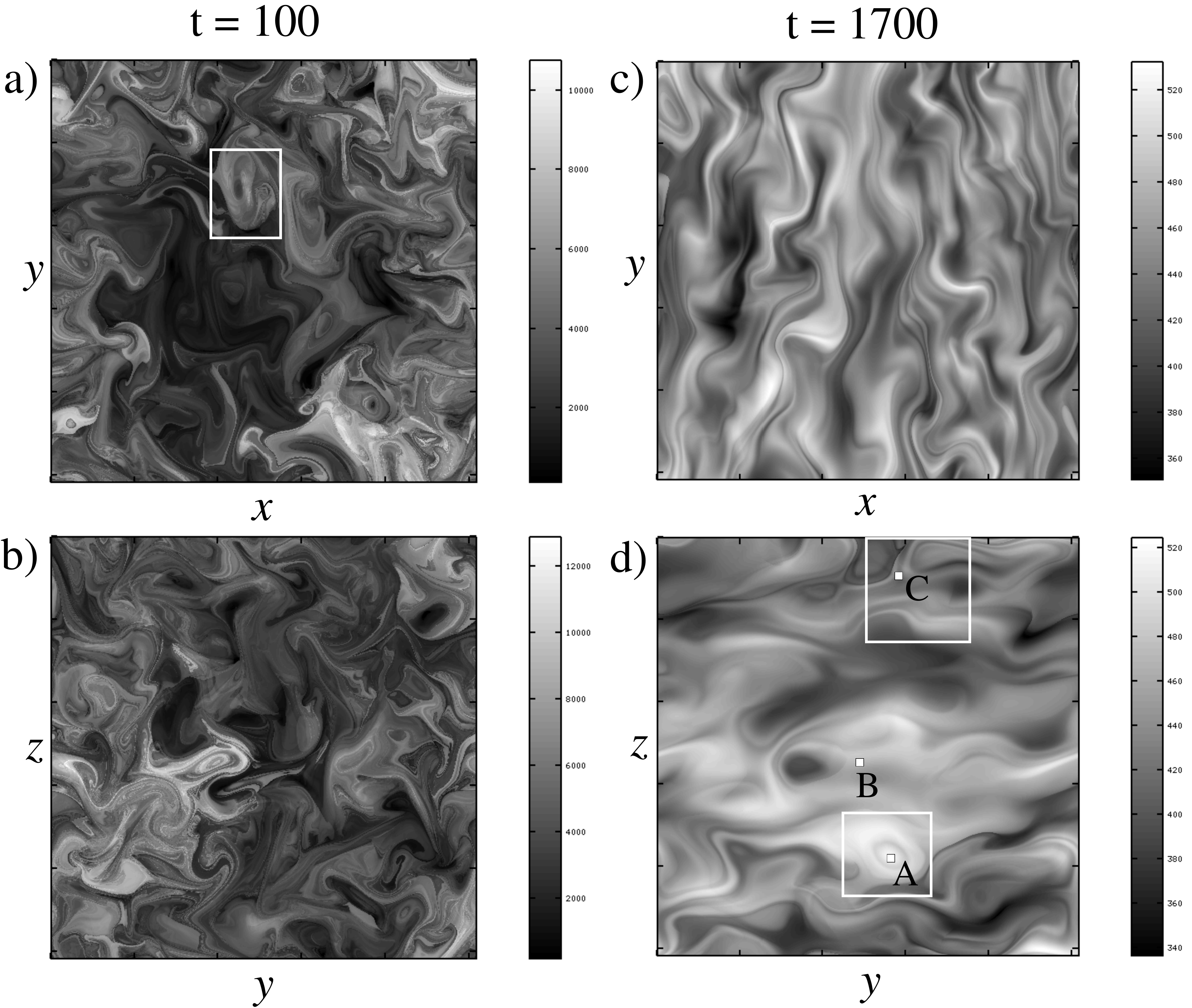}}
  \caption{Lagrangian coherent structures in the magnetic field, given by the 
           function $M$ at $t=100$ (left panel) and $t=1700$ (right panel).}
\label{fig mB}
\end{figure}

Once again, to obtain a clearer picture of magnetic coherent structures, we plot in Fig. \ref{fig mB}
the function {\it M} for $\tau=9/B_{rms}$. It is easier to spot coherent 
structures from this field, such as the one in the box in Fig. \ref{fig mB}(a). 
Function {\it M} seems to be better than the FTLE field in highlighting the
main transport barriers, filtering out spurious lines that are not so important for mixing \citep{mendoza2010}. 

\begin{figure}
  \centerline{\includegraphics[width=0.6\columnwidth]{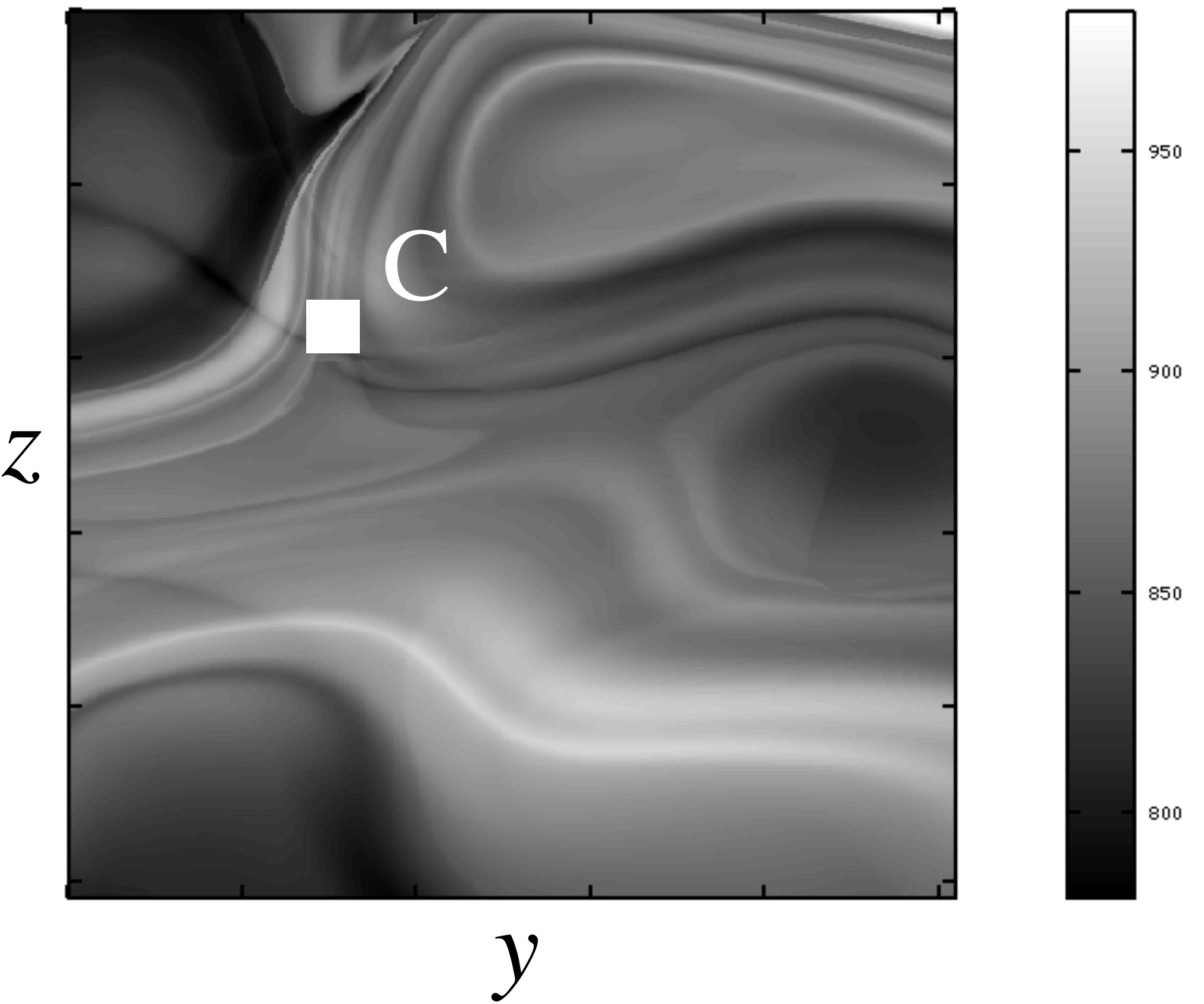}}
  \caption{Enlargement of the upper rectangle in Fig. \ref{fig mB}(d).}
\label{fig mBz}
\end{figure}

\begin{figure}
  \centerline{\includegraphics[width=1.0\columnwidth]{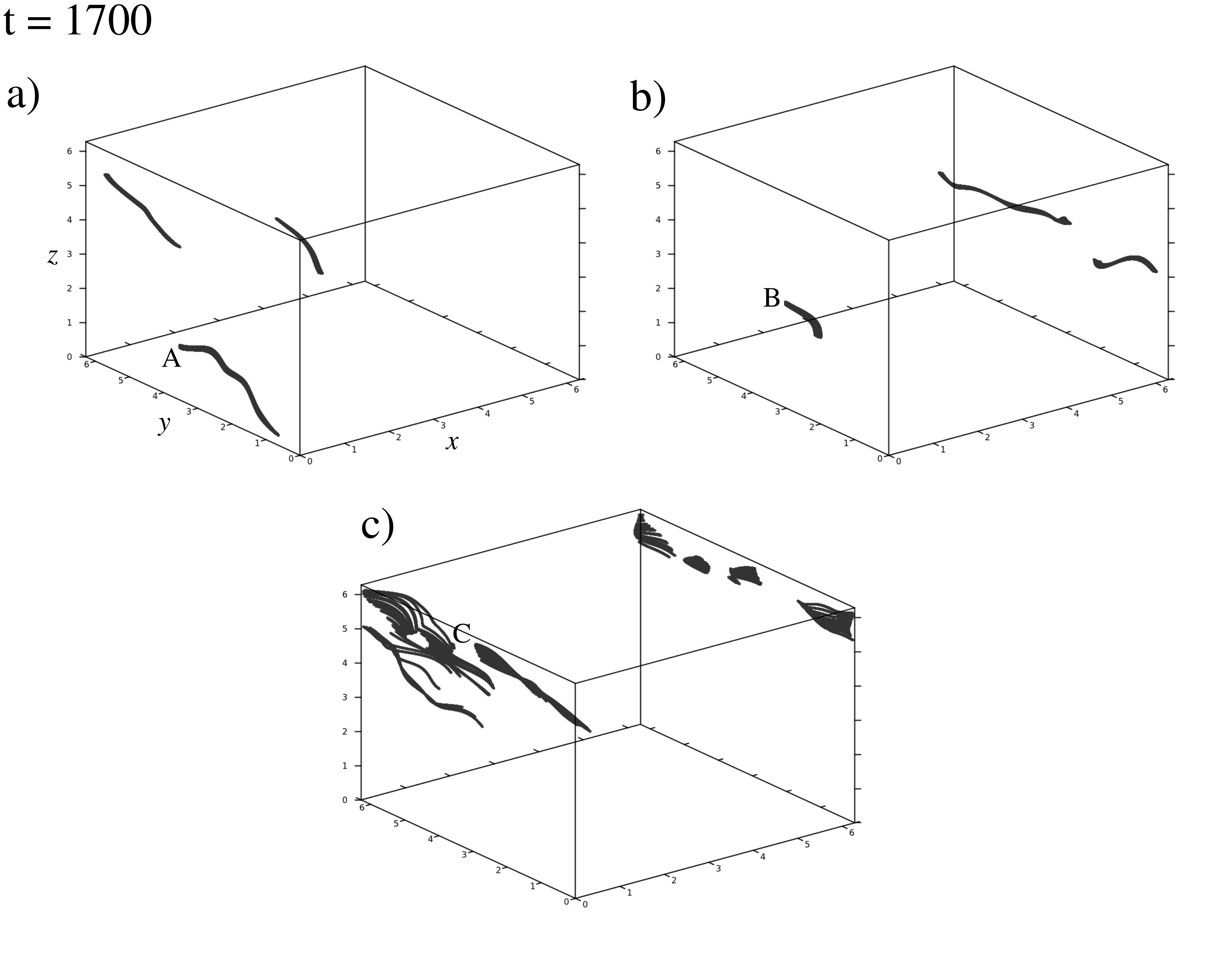}}
  \caption{Magnetic field lines produced by advecting a small blob of initial conditions in the magnetic field at $t=1700$.
(a) The initial blob is located at $A$, the point inside a magnetic vortex in Fig. \ref{fig mB}(d);
(b) the initial blob is located at $B$, the point in a smooth region of Fig. \ref{fig mB}(d); 
(c) the initial blob is located at $C$, the point at a crossing of manifolds in Fig. \ref{fig mB}(d).}
\label{fig B}
\end{figure}

As mentioned before, another feature of function {\it M} plots is that they provide both the stable and unstable
manifolds of DHTs in the same picture. In order to illustrate this feature,
three distinct regions are marked in Fig. \ref{fig mB}(d). Regions A and B are located 
in smooth parts of the {\it M} field and region C in a region where
manifolds are crossing. Smoothness of M in regions A and B
indicates that initial conditions in these regions do not perceive
nearby hyperbolic regions for $t\in(t_0-\tau, t_0+\tau)$ \citep{mendoza2010}.
An enlargement of region C is shown in Fig. \ref{fig mBz}, where 
the presence of manifolds indicates that field lines in this region 
either were dispersed in $t_0-\tau$ or will disperse in $t_0+\tau$.
We define three sets of initial conditions inside the small white squares A, B and C
in Fig. \ref{fig mB}(d),
with each square containing 25 initial conditions. 
The result of integrating Eqs. (\ref{eq line})
with each set of initial conditions for $\tau=9/B_{rms}$ field--line--time units is shown in Fig. 
\ref{fig B}. Figures \ref{fig B}(a) and (b) show the trajectories of initial conditions in regions A and B, respectively,
where it can be seen that all magnetic field lines stay close to each other, forming a magnetic
flux tube that is not dispersed in this field--line--time interval. The apparent discontinuities in field lines 
are due to the periodic boundary conditions. In Fig. \ref{fig B}(c) 
the trajectories of initial conditions at region C are shown and one can 
see that there is great chaotic dispersion of field lines due to the crossings of manifolds in this region. 

We conclude that function {\it M} can efficiently detect transport barriers and dispersion regions 
in a magnetic field. 

\section{Conclusions}\label{sec:conclusions}

Magnetohydrodynamic coherent structures have been identified in direct numerical simulations
of a nonlinear dynamo. It was shown that both Eulerian and Lagrangian tools are able to extract vortices from 
velocity and magnetic field data. Although the Eulerian tool adopted is less computationally expensive,
Lagrangian plots show finer details and can better locate the boundaries of vortices.
In addition, the Lagrangian analysis provides important information about the mixing properties
of the flow.
Regarding the numerical tools employed to detect Lagrangian coherent structures (LCSs), 
the function {\it M} seems to be less sensitive to the choice
of the integration time $\tau$ in comparison to the maximum finite--time Lyapunov exponent (FTLE).
Thus, pictures obtained with the FTLE  
can become increasingly ``noisy" with increasing $\tau$ due to the complex folding of material lines.
Although function {\it M} provides ``cleaner" pictures, the manifolds (transport barriers) are 
often not as clearly traced as in a FTLE field.
Both tools reveal the strong impact of the magnetic field on the mixing properties of
the velocity field when the system moves from the kinematic to the 
saturated dynamo phases. After the appearance of a strong mean--field,
the kinetic and magnetic coherent structures are shown to align in a preferred direction,
revealing the anisotropy developed in the vector fields.

Function {\it M} is also shown to be useful to detect manifolds of hyperbolic trajectories in the 
magnetic field, where intense transport of magnetic field lines takes place,
a feature that can be further explored to study magnetic reconnection phenomena
in plasmas.
In relation to this, Lagrangian coherent structures in photospheric velocity fields have been shown to be associated with quasi--separatrix 
layers in the magnetic field \citep{yeates2012}, which are regions of strong gradients in stretching and 
squashing of magnetic flux tubes, being identified as the preferential regions for magnetic reconnection \citep{demoulin2006,santos2008}.
Magnetic reconnection is an important phenomenon in nonlinear dynamos, since it is believed that 
it can reduce the backreaction of the Lorentz force on the velocity field \citep{blackman1996}.
Essentially, turbulent motions can cause the stretching, twisting and folding of weak magnetic
field lines in such a way as to produce the growth of magnetic flux. After the magnetic field 
reaches equipartition with the velocity field, the field lines can restrict fluid motions and 
transport of material is significantly reduced. This suppression of motions may also inhibit the
dynamo. However, if there is rapid reconnection between magnetic flux tubes, this could prevent the 
tube from backreacting. For other works on the role of magnetic reconnection in dynamo models, 
see \citet{archontis2003} and \citet{baggaley2009}.

We acknowledge Prof. R. A. Miranda, of the University of Brasilia, for his support with numerical codes.
E. L. R. acknowledges the support of FAPESP (Brazil), CNPq (Brazil) and NORDITA (Sweden). 
A. C.-L. C. acknowledges support from CNPq (Brazil), the award of a Marie Curie International
Incoming Fellowship and the hospitality of Paris Observatory. P. R. M. acknowledges the support of FAPESP (Brazil).




\end{document}